\documentclass[prb,showpacs,twocolumn]{revtex4}

\usepackage{graphicx}
\usepackage{dcolumn}
\usepackage{bm}
\usepackage{natbib}
\usepackage{amsmath}
\usepackage{here}
\usepackage{color}

\begin{document}
\preprint{preprint - not for distribution}

\title{Impact of electron heating on the equilibration between quantum Hall edge channels}

\author{Nicola Paradiso}
\author{Stefan Heun}\email{stefan.heun@nano.cnr.it}
\author{Stefano Roddaro}
\author{Lucia Sorba}
\author{Fabio Beltram}
\affiliation{NEST, Istituto Nanoscienze-CNR and Scuola Normale Superiore, Pisa, Italy}
\author{Giorgio Biasiol}
\affiliation{Istituto Officina dei Materiali CNR, Laboratorio TASC, Basovizza (TS), Italy}

\date{\today}

\begin{abstract}
When two separately contacted quantum Hall (QH) edge channels are brought into interaction, they can equilibrate their imbalance via scattering processes.
In the present work we use a tunable QH circuit to implement a junction between co-propagating edge channels whose length can be controlled with continuity. Such a variable device allows us to investigate how current-voltage characteristics evolve when the junction length $d$ is changed. Recent experiments with fixed geometry reported a significant reduction of the threshold voltage for the onset of photon emission, whose origin is still under debate. Our spatially resolved measurements reveal that this threshold shift depends on the junction length. We discuss this unexpected result on the basis of a model which demonstrates that a heating of electrons is the dominant process responsible for the observed reduction of the threshold voltage.
\end{abstract}


\maketitle

\section{Introduction}
\label{intro}

The renewed interest in \textit{integer} quantum Hall (QH) systems is principally motivated by the peculiar features of edge states.\cite{Halperin1982} They give rise to chiral one-dimensional channels that behave as perfectly collimated beams of electrons, whose trajectory,\cite{Paradiso2010} phase,\cite{Camino2005,Ji2003,Neder2006,Neder2007,Roulleau2007} back-scattering probability,\cite{Roddaro2003,Roddaro2004,Roddaro2005,Roddaro2009} and energy distribution\cite{Altimiras2010} can be accurately controlled. QH circuits are used as flexible building blocks for coherent transport devices, e.g.~the electron analogue of the Fabry-P\'erot,\cite{Camino2005} Mach-Zehnder\cite{Ji2003,Neder2006,Neder2007,Roulleau2007} or Hanbury-Brown-Twiss\cite{NederNature2007} interferometer. In recent years a number of experiments\cite{Wurtz2002,Nakajima2010,Paradiso2011} focused on a particularly promising scheme: two \textit{co-propagating} edge channels are imbalanced by means of selector gates,\cite{Komiyama1989,Komiyama1992} then brought into close proximity along a path of finite length, and are finally separated. The junction so defined allows co-propagating edges to exchange either energy and/or charge. In particular the inter-channel charge transfer allows equilibrating the initial electro-chemical potential imbalance.  The amount of scattered charge depends on the sample characteristics, on the length of the interaction path, and on the inter-channel bias.\cite{Paradiso2011}  
For small bias, the relevant equilibration process is elastic scattering induced by impurities\cite{Komiyama1989,Komiyama1992} that provide the required momentum difference between initial and final edge states. This hypothesis has been confirmed by spatially resolved measurements\cite{Paradiso2011} that related the local backscattering map to the specific impurity distribution.

For large bias, when the inter-channel imbalance exceeds the energy difference between Landau levels, also radiative transitions are observed.\cite{Ikushima2007} This effect has been recently exploited to implement an innovative converter from phase-coherent electronic states to photons in the THz region.\cite{Ikushima2010} 
While the occurrence of this radiative emission is well established, the interpretation of the threshold value is actually not clear. In fact, several papers showed\cite{Komiyama1989,Komiyama1992,Machida1996,Wurtz2002,Ikushima2010} that the threshold voltage is considerably smaller than the nominal Landau level gap $\hbar\omega_c$. Some gap reduction mechanisms have been suggested,\cite{Wurtz2002} but spectroscopic studies evidenced no deviation of the photon energy from $\hbar\omega_c$.\cite{Komiyama2006} Thus a convincing explanation for such a shift is missing so far.

In the present work we investigate how a finite imbalance is equilibrated along the junction length $d$, by studying how current-voltage characteristics change when $d$ is varied. To this end, we exploited the scanning gate microscopy technique described in Ref.~\onlinecite{Paradiso2011}. The spectral analysis reported in section~\ref{sec:expresults} reveals that the threshold voltage is \textit{lowered} when the junction length \textit{increases} and, at the same time, the transition in smoothened. In section~\ref{sec:model} we analyze the relevant inter-channel scattering processes and develop a simple model which accounts for electron heating due to hot carrier injection. The electron temperature increase produces a reduction of the threshold value due to thermal broadening of the Fermi distribution. Finally, in section~\ref{sec:discussion} we quantitatively discuss the experimental data on the basis of this model and extract the electron temperature profile along the junction.

\section{Experimental results}
\label{sec:expresults}

\begin{figure}[tbp]
\centering
\includegraphics[width=\columnwidth]{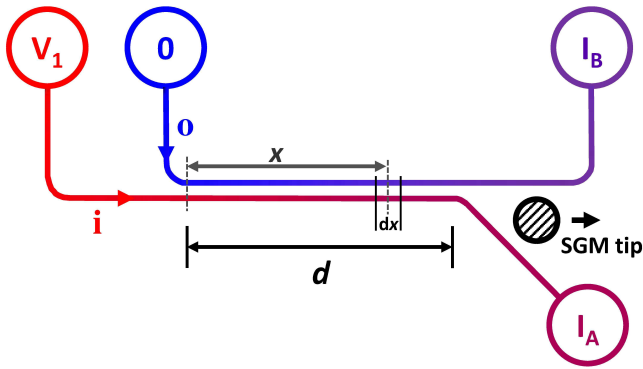}
\caption{Scheme of the experimental setup.}
\label{fig:sketch}
\end{figure}

The experimental setup is described in detail in Ref.~\onlinecite{Paradiso2011}. Devices were realized starting from a high-mobility AlGaAs/GaAs heterostructure. The 2DEG is confined 55~nm under the sample surface. By Shubnikov–-de Haas measurements we determined both the electron sheet density ($n=3.2\times 10^{15}$~m$^{-2}$) and mobility ($\mu=4.2\times 10^{2}$~m$^{2}/$V~s). A 1D channel (6~$\mu$m-long, 1~$\mu$m-wide) was defined by two Schottky gates patterned on the sample.  Measurements were performed at a base temperature of about 300~mK (electron temperature of about 400~mK) and bulk 2DES filling factor  $\nu_b=4$ at $B=3.32$~T, which corresponds to a cyclotron gap $\hbar\omega_c=5.74$~meV. Figure~\ref{fig:sketch} schematically illustrates our experiment: two cyclotron-split edge channels originate from two distinct voltage contacts at potential $V_1$ and $0$, respectively. The channels meet at the entrance of a 1D channel and travel in close proximity for a distance $d$ before they are separated by the action of the electrostatic potential induced by the biased tip of a scanning gate microscope (SGM), as shown in detail in Ref.~\onlinecite{Paradiso2011}. We label the two channels as inner ($i$) and outer ($o$) channel with chemical potential $\mu_i=eV_1$ and $\mu_o=0$, respectively.  After being separated by the SGM tip, the outgoing channels are guided to two detector contacts $I_A$ and $I_B$.

\begin{figure}[bp]
\centering
\includegraphics[width=\columnwidth]{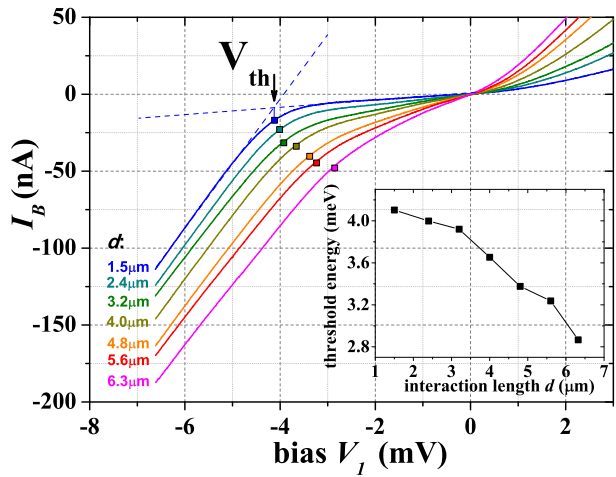}
\caption{Current-voltage characteristics for different values of the junction length $d$. The threshold points $V_{th}$ (colored dots) have been determined by extrapolating both the zero-bias and the saturation linear behavior (explicitly shown for $d=1.5$~$\mu$m), and taking the intersection point. The inset shows the dependence of the threshold voltage on $d$.}
\label{fig:IVcurves}
\end{figure}

For any value of the interaction distance $d$ compatible with device dimensions, i.e.~from 0 to 6~$\mu$m, we can measure the current-voltage ($I_B$-$V_1$) characteristics of the inter-channel charge transfer. Experimental data are shown in Fig.~\ref{fig:IVcurves}.
The first relevant feature concerns the zero-bias differential conductance which monotonically increases with the interaction length $d$. This is consistent with the differential conductance SGM plots reported in Ref.~\onlinecite{Paradiso2011}.  The curves are asymmetric around zero. While the scattered current displays a non-linear but featureless dependence on $V_1$ for positive bias,\cite{note} we will focus
on the analysis of the negative bias range ($V_1<0$, i.e.~$\mu_i>\mu_o$), where a clear transition between two distinct linear regimes occurs.
 Two linear curve sections with different slope are separated by a kink, which occurs at a certain threshold voltage $V_{th}$. We evaluate  $V_{th}$  for each individual curve by extrapolating straight lines for both the small bias and the saturation regime and taking the abscissa of the intersection point, as explicitly shown in Fig.~\ref{fig:IVcurves} for the $d=1.5$~$ \mu$m curve.
For bias smaller than $|V_{th}|$, the current-voltage characteristics are linear. The junction resistance between the two channels increases when $d$ is lowered. On the other hand, for $|V_1|>|V_{th}|$ the differential conductance saturates to $G_0\equiv e^2/h$, i.e.~half of the total conductance, so that an increase $\delta V_1$ of the input bias produces a voltage increase $\delta V_1 /2$ in both output edges.  In fact the resulting output current is $\delta I_B=G_0\delta V_1$, and therefore $\delta V_B=(h/2e^2)\delta I_B=\delta V_1/2$.
Thus, beyond the threshold, any excess of imbalance between the two edges is perfectly equilibrated. 

The most interesting feature in Fig.~\ref{fig:IVcurves} concerns the detail of the transition between the two regimes, whose position and shape clearly depends on the interaction path length $d$.  The dependence of the actual threshold voltage $|V_{th}|$ on the junction length $d$ is shown in the inset of Fig.~\ref{fig:IVcurves}. It is always smaller than $\hbar\omega_c$ and is consistently reduced by increasing $d$. At the same time the transition becomes smoother, as shown in Fig.~\ref{fig:IVcurves}. 
This is the main experimental finding of the present paper. It crucially depends on the opportunity, given by the SGM technique, to tune the junction length, keeping all the other parameters constant.

\section{Model for the inter-channel scattering}
\label{sec:model}

To discuss our model we will refer to the scheme shown in Fig.~\ref{fig:sketch}. The two edge channels meet at $x=0$ with an imbalance $\mu_i(0)-\mu_o(0)=eV_1$. 
Along the junction length $d$ the imbalance $\Delta \mu(x)\equiv \mu_i(x) - \mu_o(x) \equiv e\Delta V(x)$ will decrease due to scattering events.  In the model we assume an immediate intra-edge relaxation, so that both the chemical potential and the electron temperature $T(x)$ are well defined for each position $x$. In general, in each junction interval $dx$ the scattered current is given by
\begin{equation}
dI=\Phi(\Delta V(x), T(x))dx\label{eq:eq1},
\end{equation}   
where $\Phi$ is a general function of $\Delta V(x)$  and $T(x)$ depending on the details of the equilibration model (edge dispersion, scattering mechanisms, electron heating etc.).
The corresponding changes in the edge potentials are
\begin{eqnarray}
V_i(x+dx)&=&V_i(x)-\frac{h}{2e^2}dI \nonumber\\
V_o(x+dx)&=&V_o(x)+\frac{h}{2e^2}dI \label{eq:eq2}
\end{eqnarray}
where the factor 2 accounts for the spin degeneracy.
From equations~\ref{eq:eq1} and~\ref{eq:eq2} we obtain:
\begin{equation}
\frac{dI}{dx}=-\frac{e^2}{h}\frac{d}{dx}\Delta V(x)=\Phi(\Delta V(x),T(x))\label{eq:eq3}.
\end{equation}
The output edge currents are
\begin{eqnarray}
I_A=\frac{2e^2}{h}\frac{V_1+\Delta V(d)}{2}\nonumber\\
I_B=\frac{2e^2}{h}\frac{V_1-\Delta V(d)}{2}\label{eq:eq4},
\end{eqnarray}
whose sum equals the total input current $I_{tot}=I_A+I_B=2(e^2/h)V_1$.

\begin{figure}[tbp]
\centering
\includegraphics[width=\columnwidth]{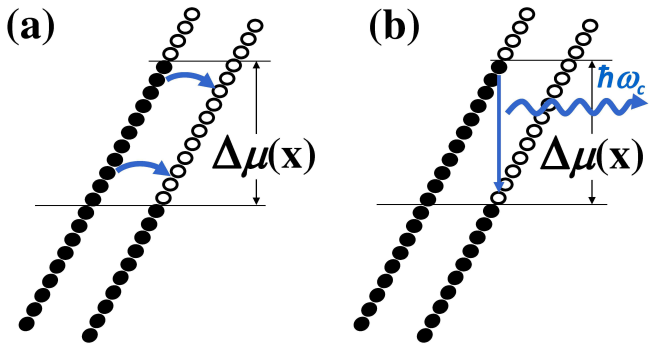}
\caption{(a) Scheme of the impurity-induced elastic scattering for a non-interacting electron system. (b) When the chemical potential of the inner edge becomes higher than the outer one by at least the cyclotron gap $\hbar\omega_c$, vertical radiative transitions can occur. Notice that for opposite polarity vertical transitions are suppressed.}
\label{fig:elastic}
\end{figure}

Inter-channel scattering can originate from several processes. For low bias, the inter-edge electron transfer can be either induced by impurity or phonon scattering.\cite{Komiyama1989,Komiyama1992} The latter, however, was shown\cite{Komiyama1989,Komiyama1992} to be less important when the base temperature is smaller than 1~K. The relevant process (sketched in Fig.~\ref{fig:elastic}(a)) is thus the elastic scattering induced by sharp impurity potentials which provide the change in momentum needed for the inter-channel transition. 
The infinitesimal scattered current in the interval $dx$ is
\begin{equation}
dI=\int^{\infty}_{-\infty}eD(\epsilon)\mathcal{T}(\epsilon)(f_{\mu_i,T}(\epsilon)-f_{\mu_o,T}(\epsilon))d\epsilon\label{eq:eq5},
\end{equation}
where $D(\epsilon)$ is the density of states around the energy $\epsilon$ and $\mathcal{T}(\epsilon)$ is the elastic scattering probability per unit time.

 In order to estimate expressions as the one on the right hand side of Eq.~\ref{eq:eq5}, a model for the edge dispersion is needed. In this paper we will assume the simplest case, i.e.~a linear dispersion, a choice that will be justified in Section~\ref{sec:discussion} on the basis of the observed temperature effects. 
In this approximation, we can assume both $D$ and $\mathcal{T}$ as constant in the energy window $e\Delta V$. In this case the density of states is $D(\epsilon)=2dx/(hv_d)$, where $v_d$ is the drift velocity. Thus (see appendix~\ref{sec:AppA})
\begin{eqnarray}
dI &=& dx\frac{2e\mathcal{T}_0}{hv_d}\int^{\infty}_{-\infty}(f_{\mu_i,T}(\epsilon)-f_{\mu_o,T}(\epsilon))d\epsilon\nonumber\\
   &=& dx\frac{2e^2\mathcal{T}_0}{hv_d} \Delta V(x) \label{eq:eq6},
\end{eqnarray}
where $\mathcal{T}_0$ is the constant transmission probability. For this process  $\Phi$ is linear in $\Delta V(x)$  and does not depend on $T$. $I_B$-$V_1$ curves can thus be calculated by solving the ordinary differential equation~\ref{eq:eq3} for $\Delta V(x)$ with boundary condition $\Delta V(0)=V_1$, which gives an exponential decay of the edge imbalance
\begin{equation}
\Delta V(x) = V_1  e^{-\frac{2\mathcal{T}_0}{v_d}x}\label{eq:eq7}.           
\end{equation}
This exponential behavior was assumed in literature\cite{Komiyama1989,Komiyama1992,Muller1992} to describe the zero-bias inter-channel scattering in the limit of a  uniform distribution of scattering centers. The characteristic length in this case is $\ell_{eq}=v_d/(2\mathcal{T}_0)$, i.e.~the average distance between two scattering events. We experimentally verified this exponential decay in our previous work.\cite{Paradiso2011}
Furthermore, the output current $I_B$ is linear in $V_1$ (ohmic behavior):
\begin{equation}
I_B =\frac{2e^2}{h}\frac{V_1-\Delta V(d)}{2}=V_1 \frac{2e^2}{h}\frac{1-e^{-\frac{d}{\ell_{eq}}}   }{2}.\label{eq:eq8}
\end{equation}

At higher imbalance, comparable to the Landau level gap $\hbar\omega_c$, other equilibration processes become possible. When $\mu_i>\mu_o$ vertical radiative transitions from the inner edge to the outer one are enabled, as depicted in Fig.~\ref{fig:elastic}(b). Non-vertical relaxation could in principle occur via phonon-assisted transitions. However, this is a second-order effect that can in first approximation be disregarded, at least for low temperatures.
The infinitesimal scattered current due to vertical transitions is then given by
\begin{equation}
dI=\int^{\infty}_{-\infty}eD(\epsilon)\mathcal{T}_{1}(\epsilon)[f_{\mu_i,T}(\epsilon)(1-f_{\mu_o,T}(\epsilon-\hbar\omega_c))]d\epsilon\label{eq:eq9},
\end{equation}
where $\mathcal{T}_1$ is the probability per unit time for the transition $\epsilon \rightarrow\epsilon-\hbar\omega_c$. Since the Landau level bands are parallel, the transition probability is constant in energy. Therefore we can simplify Eq.~\ref{eq:eq9} 
\begin{eqnarray}
dI&=&dx\frac{2e\mathcal{T}_1}{hv_d}\int^{\infty}_{-\infty}[f_{\mu_i,T}(\epsilon)(1-f_{\mu_o,T}(\epsilon-\hbar\omega_c))]d\epsilon\nonumber\\
&=&dx\frac{2e\mathcal{T}_1}{hv_d}\left( \frac{e\Delta V(x)-\hbar\omega_c}{1-e^{\frac{\hbar\omega_c-e\Delta V(x)} {k_BT(x)}}}\right)
\label{eq:eq10}
\end{eqnarray}
where the integration is explicitly shown in appendix~\ref{sec:AppA}. In the  $\Phi$ function we also have  a non-linear addendum, thus the integration of Eq.~\ref{eq:eq3} has to be performed numerically. At low temperature, due to the exponential term, the effect of the term in Eq.~\ref{eq:eq10} is negligible for $\Delta V (x)$ below the threshold $ \hbar \omega_c$. For $\Delta V (x)> \hbar \omega_c$ the availability of empty states in the lower Landau level gives rise to a strong radiative relaxation. As shown in recent experiments,\cite{Ikushima2010} the photons emitted in this process can be collected with a suitable waveguide and detected.

So far we completely neglected the effect of the electron heating due to the injection of hot carriers. In order to obtain a quantitative estimate of the amount of energy transferred to the electron system, we need to first estimate the total energy increase of an edge channel when we increase its chemical potential from the ground level $\mu=\mu_0$ to $\mu=\mu_j$ and its temperature from $T=0$ to $T=T_j$
\begin{eqnarray}
\mathcal{E}_j
&=&\int^{\infty}_{-\infty}\frac{2d}{hv_d}(\epsilon-\mu_0)(f_{\mu_j,T}(\epsilon)-f_{\mu_0,0}(\epsilon))d\epsilon\nonumber\\
&\approx&\frac{1}{2}\frac{2\tau}{h}(\mu_j-\mu_0)^2+\frac{2\tau}{h}\frac{\pi^2}{6}k_B^2T_j^2
 \label{eq:eq11},
\end{eqnarray}
where in the second line we approximated the integral with the first order Sommerfeld expansion (as shown in detail in appendix~\ref{sec:AppB}) and $\tau\equiv d/v_d$.

To calculate explicitly the output temperature $T(x)$ we will assume energy conservation in each infinitesimal element $dx$
\begin{equation}
\mathcal{E}_i(x)+\mathcal{E}_o(x)=\mathcal{E}_i(x+dx)+\mathcal{E}_o(x+dx)\label{eq:eq12}
\end{equation}
together with three additional approximations: (i) the two edges immediately restore the thermal equilibrium after each scattering event; (ii) the temperature is approximately the same in both edges $T_i(x)=T_o(x)=T(x)$, with $T(0)=T_{in}$, where $T_{in}$ is the bulk electron temperature; (iii) in each element $dx$ only the ohmic part of the scattered current $dI$ contributes to the electron heating. In fact, only the elastic process transfers hot carriers between the two edges, while the radiative term allows electrons to relax by photon emission.
With these assumptions, after substituting the expression in Eq.~\ref{eq:eq11} into Eq.~\ref{eq:eq12} as shown in Appendix~\ref{sec:AppC}, we can deduce an equation which relates the change in temperature with the local imbalance  
\begin{equation}
\frac{d}{dx}T(x)=\frac{3e^2}{4\pi^2k_B^2\ell_{eq}}\frac{\Delta V^2(x)}{T(x)}\label{eq:eq13}.
\end{equation}
Thus Eq.~\ref{eq:eq3} must be solved together with Eq.~\ref{eq:eq13} to extract both $T(x)$ and $V(x)$. Due to the electron heating, the onset of radiative transitions is shifted below the cyclotron gap value $\hbar \omega_c$ since thermally excited electrons leave available states in a range of about $k_BT$ around the chemical potential of the lower level. The transition itself becomes smoother, since the expression in Eq.~\ref{eq:eq10} is less steep at higher temperatures.

\section{Discussion}
\label{sec:discussion}

\begin{figure}[bp]
\centering
\includegraphics[width=\columnwidth]{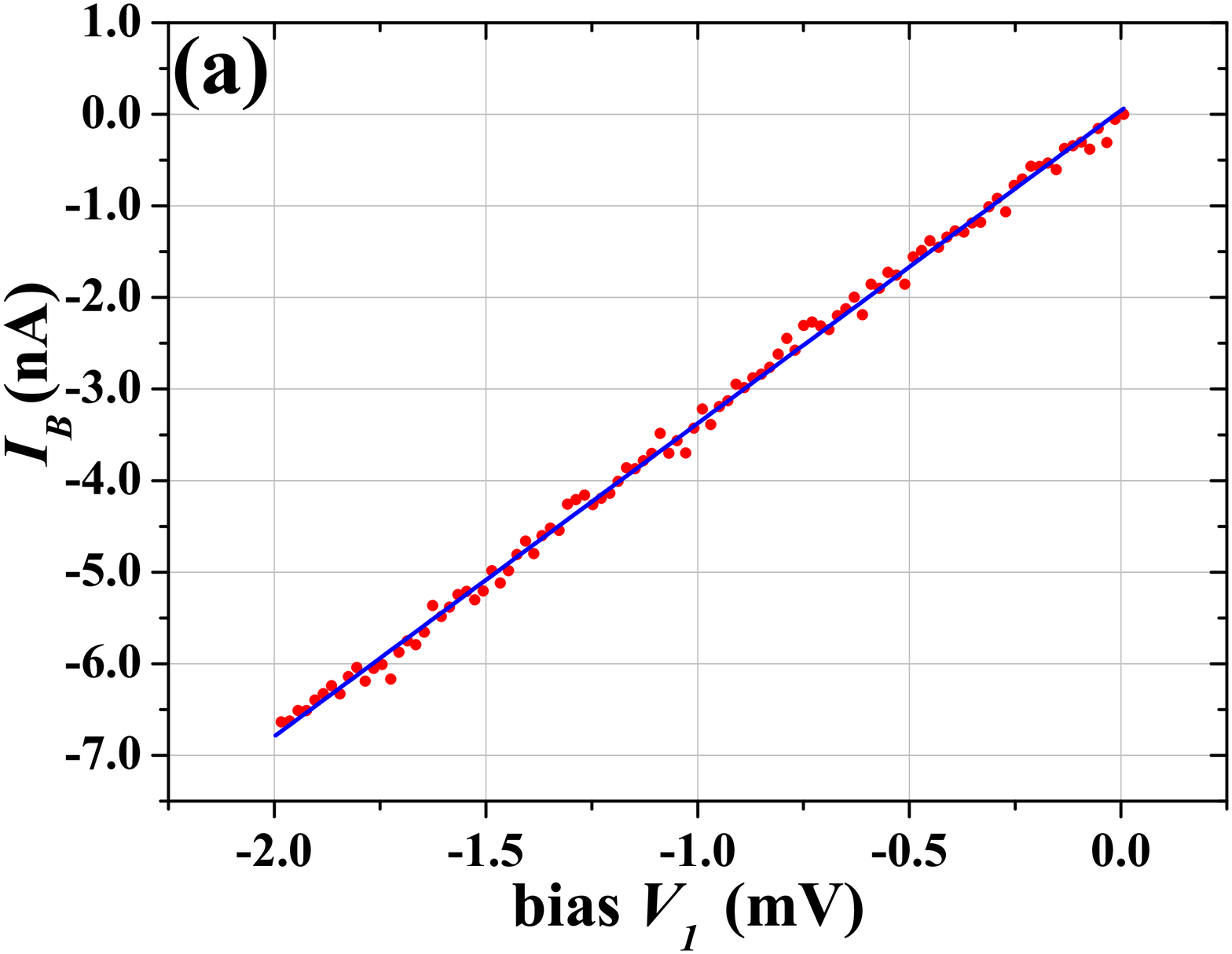}
\includegraphics[width=\columnwidth]{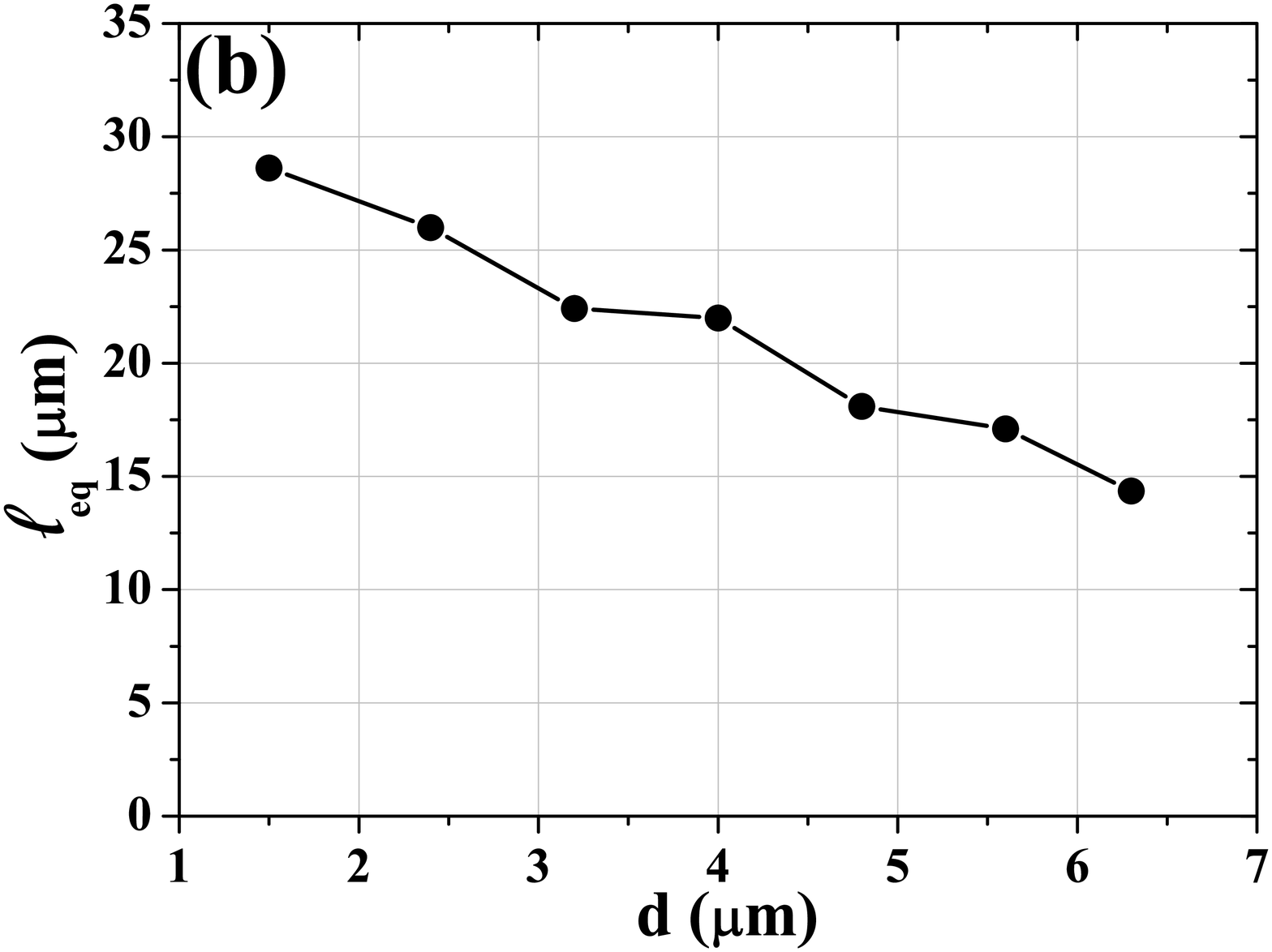}
\caption{(a) Detail of the $I_B$-$V_1$ characteristics in the range -2~mV$<V_1<0$, for $d=2.4$~$\mu$m (red dots). The behavior is ohmic as evidenced by the linear fit (blue line). (b) Plot of $\ell_{eq}$ for different junction lengths $d$.}
\label{fig:linear}
\end{figure}

Figure~\ref{fig:linear}(a) shows the $I_B$-$V_1$ characteristics (red dots) for the  $d=2.4$~$\mu$m case. The behavior is clearly ohmic, as confirmed by a linear fit (blue line, adjusted $R^2=0.997$). This agrees with the predictions of our model for low bias, when radiative emission is negligible and Eq.~\ref{eq:eq8} applies. The zero-bias differential conductance depends on the distribution of scattering centers inside the constriction. Equation~\ref{eq:eq8} allows us to obtain the equilibration length $\ell_{eq}$ by fitting the $I_B$-$V_1$ curves in the linear region. Figure~\ref{fig:linear}(b) displays the different $\ell_{eq}$ values obtained for each junction length $d$. The average $\ell_{eq}$ value (21~$\mu$m) is consistent with the one reported in Ref.~\onlinecite{Paradiso2011} (15~$\mu$m), considering that those results were obtained from different samples. The graph evidences that $\ell_{eq}$ depends on $d$. 
As shown in Ref.~\onlinecite{Paradiso2011}, the actual impurity density is highly sample-dependent and can fluctuate along the  inter-channel junction. The monotonical decrease observed in Fig.~\ref{fig:linear}(b) could however indicate that for short $d$ the scattering centers are somewhat less effective, due to the fact that the edges are smoothly brought into interaction and separated. Therefore the inter-channel separation is larger at the constriction ends than at the inner points. These boundary effects are more important for smaller $d$. 

\begin{figure}[tbp]
\centering
\includegraphics[width=\columnwidth]{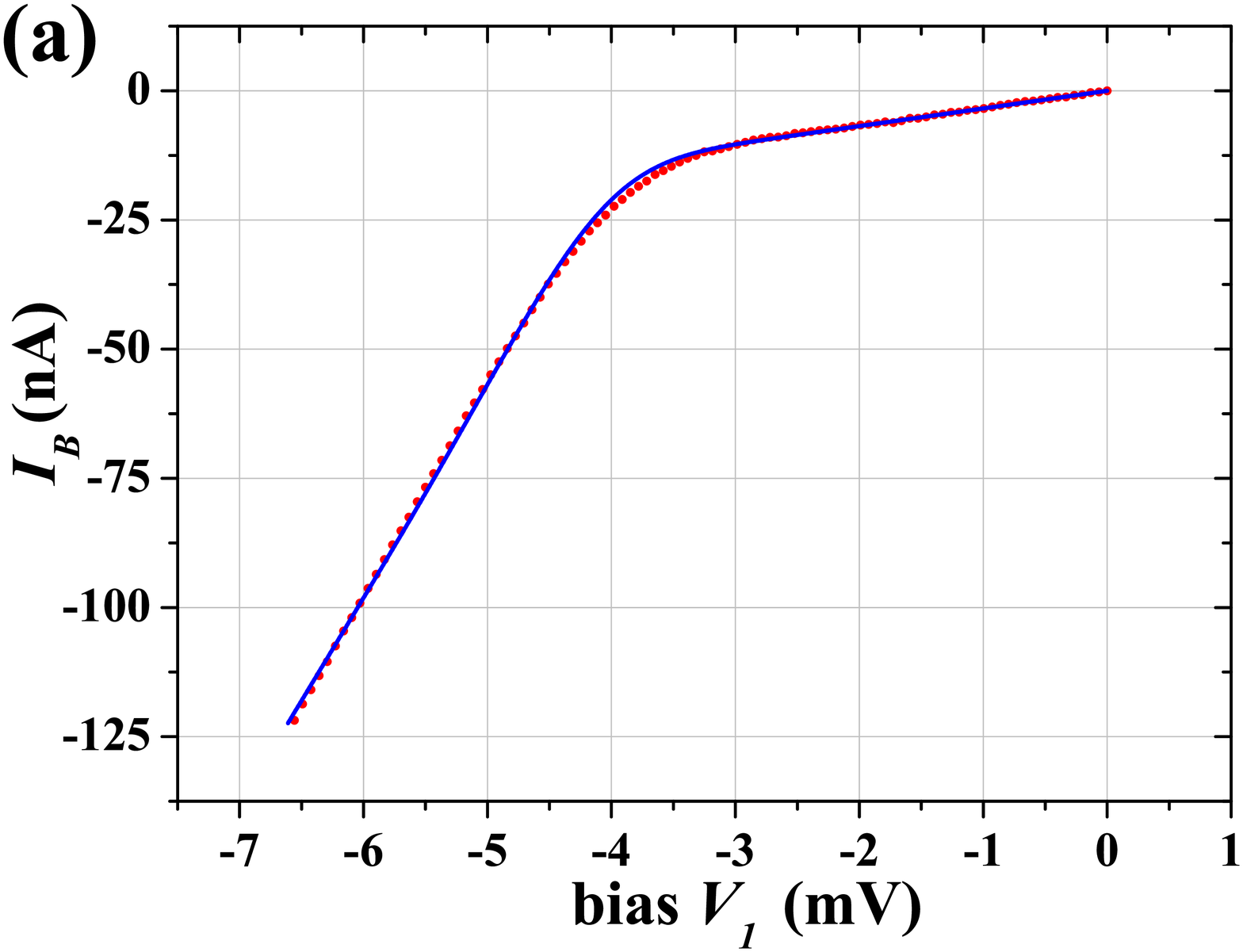}
\includegraphics[width=\columnwidth]{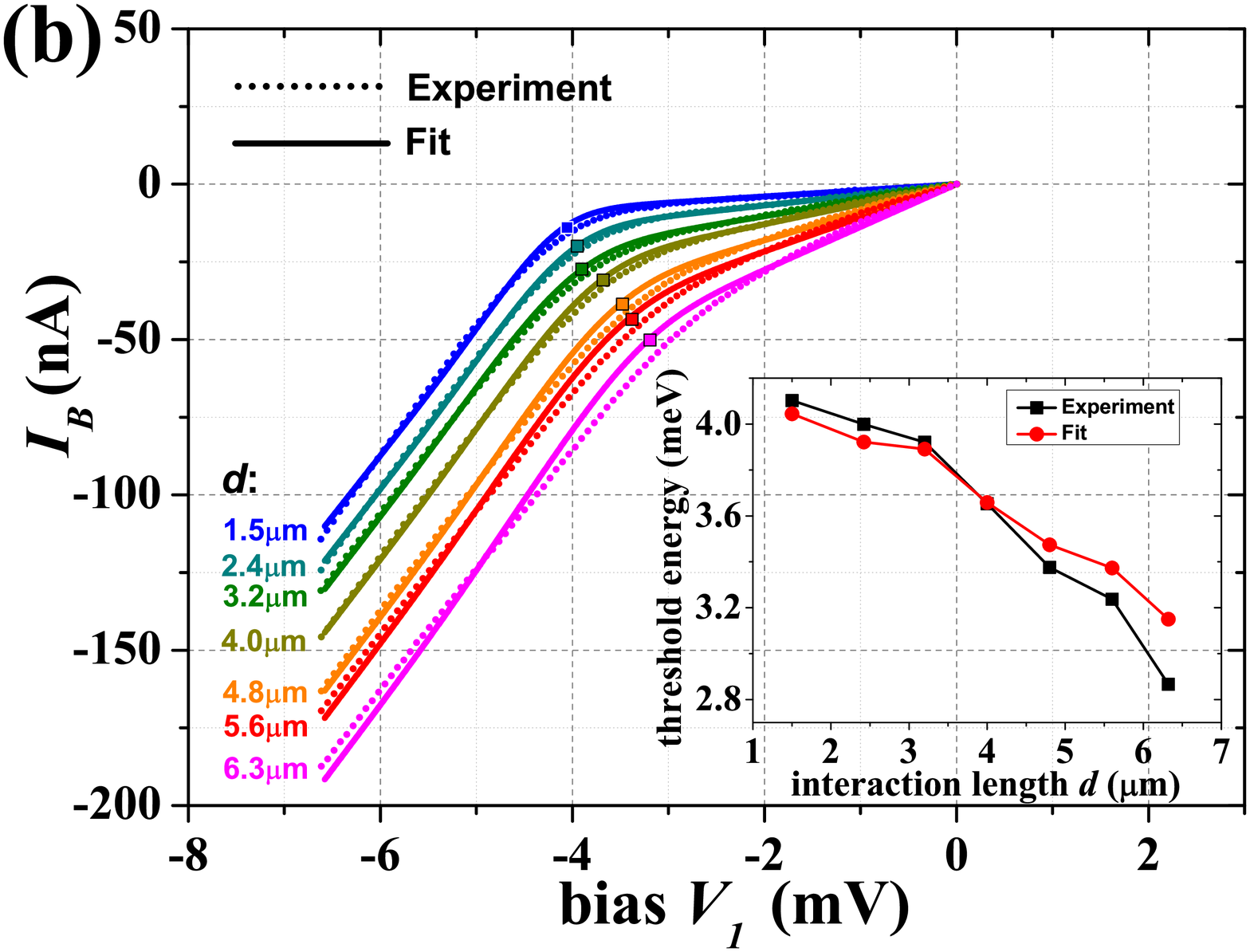}
\caption{(a) Fit (blue line) of the $I_B$-$V_1$ curve for $d=2.4$~$\mu$m (red dots) using solutions of Eqs.~\ref{eq:eq3} and~\ref{eq:eq13}, with the parameter $\ell_{eq}$ obtained from the previous linear fits. (b) Complete set of fitting curves (solid lines) for all measured $d$ values, together with the corresponding experimental data of Fig.~\ref{fig:IVcurves} (dotted lines).
(Inset) Threshold voltages plotted as a function of $d$ as deduced from the fitting curves (red dots), together with the values directly extracted from Fig.~\ref{fig:IVcurves} (black squares).
}
\label{fig:fit}
\end{figure}

The previous results provide the first of the two free parameters of our model, namely $\ell_{eq}$ and $\mathcal{T}_1$. Therefore we fit the experimental curves in Fig.~\ref{fig:IVcurves} with the functions obtained solving Eqs.~\ref{eq:eq3} and~\ref{eq:eq13}, with the only free parameter $\mathcal{T}_1$. The fit for $d=2.4$~$\mu$m is displayed in Fig.~\ref{fig:fit}(a), together with the experimental data. The agreement between the two curves is remarkable: our simple model reproduces both the shift and the smoothing at the threshold, i.e.~the two main features observed in Fig.~\ref{fig:IVcurves}. The threshold shift can be better seen in Fig.~\ref{fig:fit}(b), where we plot the fits for all experimental curves of Fig.~\ref{fig:IVcurves} (solid lines), together with the corresponding experimental data (dotted lines). In the inset we show a comparison between the threshold voltage values extracted from the fitting curves and the ones directly estimated from the $I_B$-$V_1$ characteristics. This graph clearly indicates that the present model suitably describes the observed threshold reduction. The value for the Landau level gap ($\hbar\omega_c=5.74$~meV) was kept constant in these fits. This value turns out to be optimal once both $\ell_{eq}$ and $\mathcal{T}_1$ have been determined, because then a further adjustment of the gap only decreases the fit quality. 

This result explains the reduction of the threshold for photon emission observed in several experiments.\cite{Ikushima2010,Wurtz2002} The significant deviation from $\hbar\omega_c/e$ is an effect due to the electron heating induced by the injection of hot carriers in the outer edge via elastic scattering. 

\begin{figure}[tbp]
\centering
\includegraphics[width=\columnwidth]{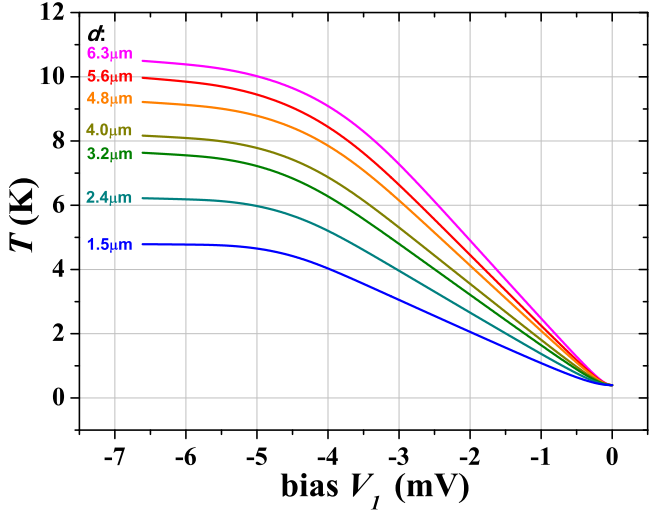}
\caption{Bias dependence of the  outgoing electron temperature plotted for different $d$ values. The curves are obtained from Eq.~\ref{eq:eq13}, with the initial condition $T(0)=400$~mK. The parameters $\ell_{eq}$ and $\mathcal{T}_1$ are given by the previous fits of the experimental data.   }
\label{fig:temp}
\end{figure}

To quantitatively estimate the electron temperature increase, we solved Eq.~\ref{eq:eq13}, using the parameters $\ell_{eq}$ and $\mathcal{T}_1$ provided by the previous fits, with the initial condition $T_{in}=400$~mK. Figure~\ref{fig:temp} shows the solutions for the $d$ values corresponding to the experimental data in Fig.~\ref{fig:IVcurves}.
For very small bias the temperature increases almost quadratically with the imbalance, while for intermediate values the behavior is approximatively linear, with a slope proportional to $d$. Finally, the temperature tends to saturate at the onset of radiative emission, which suppresses further injection of hot electrons into the outer edge. At saturation, the output edge temperatures are by far larger than the base temperature. 

In section~\ref{sec:model} we considered a linear edge dispersion, which neglects effects of edge reconstruction due to electron-electron interaction.\cite{Chklovskii1992}  We have also developed alternative models, which take into account the effect of the compressible and incompressible stripes at the sample edge. While such more complex analysis correctly predicts the linear behavior at low bias, it is less satisfactory in describing the threshold evolution, although it contains more adjustable parameters (as the compressible and incompressible stripe widths). We interpreted such discrepancy as the effect of the high electron temperature induced by the elastic scattering processes and present on most part of the edge junction. As edge reconstruction is known to be quickly washed out by temperature,\cite{Machida1996,Lier1994,Suzuki1993} we have therefore rather chosen a simple model with a linear edge  dispersion, which indeed captures the relevant features observed in the experiment.


\section{Conclusion}
\label{sec:conclusion}
We demonstrated a tunable-length junction between highly imbalanced edge channels in the quantum Hall regime. The measurements of its current-voltage characteristics clearly evidence that the threshold voltage for the onset of radiative emission depends on the junction length $d$. We show how this  behavior can be explained by a simple model accounting for the heating effect due to the elastic scattering of hot carriers.

\begin{acknowledgments}
We acknowledge financial support from the Italian Ministry of
Research (MIUR-FIRB projects RBIN045MNB, RBIN048ABS, RBID08B3FM, RBIN067A39\_002, and RBIN06JB4C).
\end{acknowledgments} 

\appendix{\section{Integration of expressions containing Fermi functions}
\label{sec:AppA}
In Eq.~\ref{eq:eq6} we evaluated the integral
\begin{eqnarray}
\int^{\infty}_{-\infty}(f_{\mu_i,T}(\epsilon)-f_{\mu_o,T}(\epsilon))d\epsilon=\nonumber\\
=\int^{\infty}_{-\infty} \left(\frac{1}{1+e^{\frac{\epsilon-\mu_i}{k_BT}}}-\frac{1}{1+e^{\frac{\epsilon-\mu_o}{k_BT}}}\right) d\epsilon \label{eq:eqa1}.
\end{eqnarray}
Defining $x\equiv\epsilon/k_BT$, $x_i\equiv\mu_i/k_BT$ and $x_o\equiv\mu_o/k_BT$, we have
\begin{equation}
\int^{\infty}_{-\infty} \left(\frac{1}{1+e^{x-x_i}}-\frac{1}{1+e^{x-x_o}}\right) k_BT dx \label{eq:eqa2}.
\end{equation}
A primitive of the expression in brackets is
\begin{equation}-\ln(1+e^{x-x_i})+\ln(1+e^{x-x_o})\label{eq:eqa3},
\end{equation}
thus
\begin{eqnarray}
& &\int^{\infty}_{-\infty} \left(\frac{1}{1+e^{x-x_i}}-\frac{1}{1+e^{x-x_o}}\right) k_BT dx = \nonumber\\
&=&  \lim_{x \to +\infty} k_BT \left[ -\ln(1+e^{x-x_i})+\ln(1+e^{x-x_o})    \right]+\nonumber\\   
&-&\lim_{x \to -\infty} k_BT \left[  -\ln(1+e^{x-x_i})+\ln(1+e^{x-x_o}) \right] =\nonumber\\  
&=&k_BT(x_i-x_o)-0=\mu_i-\mu_o=e\Delta V \label{eq:eqa4}.
\end{eqnarray}

In Eq.~\ref{eq:eq10} we evaluated the integral
\begin{equation}
\int^{\infty}_{-\infty}[f_{\mu_i,T}(\epsilon)(1-f_{\mu_o,T}(\epsilon-\hbar\omega_c))]d\epsilon. \label{eq:eqa5}
\end{equation}
Defining $x\equiv\epsilon/k_BT$, $x_i\equiv\mu_i/k_BT$ and $x_o\equiv (\hbar\omega_c+\mu_o)/k_BT$, we have
\begin{equation}
\int^{\infty}_{-\infty} \left[\frac{1}{1+e^{x-x_i}}\left(1-\frac{1}{1+e^{x-x_o}}\right)\right] k_BT dx. \label{eq:eqa6}
\end{equation}
A primitive of the expression in square brackets is
\begin{equation}
\frac{e^{x_i}}{e^{x_i}-e^{x_o}}\ln\left(  \frac{e^{x_o}+e^{x}}{e^{x_i}+e^{x}}    \right)
\label{eq:eqa7},
\end{equation}
thus
\begin{eqnarray}
& &\int^{\infty}_{-\infty} \left[\frac{1}{1+e^{x-x_i}}\left(1-\frac{1}{1+e^{x-x_o}}\right)\right] k_BT dx = \nonumber\\
&=&  \lim_{x \to +\infty} k_BT \left[ \frac{e^{x_i}}{e^{x_i}-e^{x_o}}\ln\left(  \frac{e^{x_o}+e^{x}}{e^{x_i}+e^{x}}    \right)   \right]+\nonumber\\   
&-&\lim_{x \to -\infty} k_BT \left[ \frac{e^{x_i}}{e^{x_i}-e^{x_o}}\ln\left(  \frac{e^{x_o}+e^{x}}{e^{x_i}+e^{x}}    \right) \right] =\nonumber\\  
&=&k_BT \left[0-\frac{e^{x_i}}{e^{x_i}-e^{x_o}}(x_o-x_i) \right]=\nonumber\\
&=&k_BT \frac{x_i-x_o}{1-e^{x_o-x_i}}=\frac{e\Delta V-\hbar\omega_c}{1-e^{\frac{\hbar\omega_c-e\Delta V}{k_BT}}} \label{eq:eqa8}.
\end{eqnarray}

\section{First order approximation to the edge energy}
\label{sec:AppB}
In order to evaluate the first line of Eq.~\ref{eq:eq11} we exploit the Sommerfeld expansion
\begin{eqnarray}
& &\int^{\infty}_{-\infty}\frac{g(\epsilon)}{1+e^{\frac{\epsilon-\mu}{k_BT}}}d\epsilon=\nonumber\\
&=&\int^{\mu}_{-\infty}g(\epsilon)d\epsilon+\frac{\pi^2}{6}k_B^2T^2g^{\prime} (\mu)+O\left( \frac{k_BT}{\mu} \right)^4 \label{eq:eqb1}
\end{eqnarray}
where $g(\epsilon)$ is a generic function of $\epsilon$ and $g^{\prime} (\mu)$ is its first derivative evaluated at $\epsilon=\mu$. By applying this relation to Eq.~\ref{eq:eq11} we obtain
\begin{eqnarray}
&  &\int^{\infty}_{-\infty}\frac{2d}{hv_d}\frac{\epsilon-\mu_0}{1+e^{\frac{\epsilon-\mu}{k_BT}}}d\epsilon-\int^{\mu_0}_{-\infty}\frac{2d}{hv_d}(\epsilon-\mu_0)d\epsilon \approx \nonumber\\
&\approx&\int^{\mu_j}_{\mu_0}\frac{2d}{hv_d}(\epsilon-\mu_0)d\epsilon +\frac{2d}{hv_d}\frac{\pi^2}{6}k_B^2T_j^2=\nonumber\\
&=&\frac{1}{2}\left( \frac{2\tau}{h} \right) (\mu_j-\mu_0)^2+\left(\frac{2\tau}{h}\right)\frac{\pi^2}{6}k_B^2T_j^2. \label{eq:eqb2}
\end{eqnarray}

\section{Determination of $T(x)$}
\label{sec:AppC}
When the electron temperature is non-zero, the expression for the total edge energy has an extra term proportional to $T^2$, as seen in Eq.~\ref{eq:eqb2}. We can thus define the electrostatic and the thermal component of the total edge energy:
\begin{eqnarray} 
E^{el} &\equiv&  \frac{1}{2}\left( \frac{2\tau}{h} \right) (\mu_j-\mu_0)^2=\frac{1}{2}\left( \frac{2\tau}{h} \right) e^2V_j^2\nonumber\\
E^{th} &\equiv&  \left(\frac{2\tau}{h}\right)\frac{\pi^2}{6}k_B^2T_j^2\label{eq:eqc1}
\end{eqnarray}
where $V_j$ is the edge voltage referred to the ground.

Equation~\ref{eq:eqc1} allows us to evaluate Eq.~\ref{eq:eq12}.  As already explained in the paper, only  elastic scattering processes transfer hot carriers between the edges, while the radiative term allows electrons to relax by photon emission. Thus we modify Eqs.~\ref{eq:eq2} as follows
\begin{eqnarray}
V_i(x+dx)&=&V_i(x)-\frac{h}{2e^2}dI^{elast.}\nonumber\\
         &=&V_i(x)-\frac{h}{2e^2}\frac{e^2}{h}\frac{1}{\ell_{eq}}\Delta V(x)dx\nonumber\\
V_o(x+dx)&=&V_o(x)+\frac{h}{2e^2}dI^{elast.} \nonumber\\
         &=&V_o(x)+\frac{h}{2e^2}\frac{e^2}{h}\frac{1}{\ell_{eq}}\Delta V(x)dx.
         \label{eq:eqc2}
\end{eqnarray}

After evaluating Eq.~\ref{eq:eq12} with Eq.~\ref{eq:eqc1}, using the substitutions~\ref{eq:eqc2} we obtain
\begin{equation}
\frac{2\pi^2}{3}k_B^2T(x) dT = \frac{e^2}{2}\frac{1}{\ell_{eq}}\Delta V^2(x)dx
\label{eq:eqc3}
\end{equation}
(where $T_i(x)=T_o(x)=T(x)$) from which Eq.~\ref{eq:eq12} easily follows.\vfill
}

\end{document}